\documentclass[fleqn,usenatbib]{mnras}

\usepackage{newtxtext,newtxmath}
\usepackage{ulem}

\usepackage[T1]{fontenc}

\DeclareRobustCommand{\VAN}[3]{#2}
\let\VANthebibliography\thebibliography
\def\thebibliography{\DeclareRobustCommand{\VAN}[3]{##3}\VANthebibliography}

\usepackage{graphicx}	
\usepackage{amsmath}	
\usepackage{xltabular}
\usepackage{xtab,booktabs}
\usepackage{xspace}
\usepackage{longtable}

\newcommand{\hii}{H\thinspace\textsc{ii}\xspace}
\newcommand{\divingTD}{DIVING$^\mathrm{3D}$\xspace}

\title[The \divingTD Survey I]{The \divingTD Survey -- Deep IFS View of Nuclei of Galaxies -- I. Definition and Sample Presentation}

\author[J. E. Steiner et al.]{J. E. Steiner$^{1}$\thanks{Deceased}, R. B. Menezes$^{2}$\thanks{\href{mailto:roberto.menezes@maua.br}{roberto.menezes@maua.br}},
T. V. Ricci$^3$\thanks{\href{mailto:tiago.ricci@uffs.edu.br}{tiago.ricci@uffs.edu.br}},
Patr\'icia da Silva$^{1}$\thanks{\href{mailto:p.silva2201@gmail.com}{p.silva2201@gmail.com}},
R. Cid Fernandes$^{4}$,
N. Vale Asari$^{4,5}$\thanks{Royal Society--Newton Advanced Fellowship},
\newauthor{M. S. Carvalho$^{4}$,
D. May$^1$, Paula R. T. Coelho$^{1}$ and A.~L. de Amorim$^{4}$}
\\
$^{1}$Instituto de Astronomia, Geof\'isica e Ci\^encias Atmosf\'ericas, Departamento de Astronomia, Universidade de S\~ao Paulo, 05508-090, SP, Brazil\\
$^{2}$Instituto Mau\'a de Tecnologia, Pra\c{c}a Mau\'a 1, 09580-900, S\~ao Caetano do Sul, SP, Brazil\\
$^{3}$Universidade Federal da Fronteira Sul, Cerro Largo, 97900-000, RS, Brazil\\
$^{4}$Universidade Federal de Santa Catarina, Departamento de F\'isica--CFM, C.P. 476, 88040-900, Florian\'opolis, SC, Brazil\\
$^{5}$University of St Andrews, School of Physics and Astronomy, North Haugh, St Andrews KY16 9SS, UK}
\date{Accepted XXX. Received YYY; in original form ZZZ}

\pubyear{2021}

\begin{document}
\label{firstpage}
\pagerange{\pageref{firstpage}--\pageref{lastpage}}
\maketitle

\begin{abstract}
We present the Deep Integral Field Spectrograph View of Nuclei of Galaxies (\divingTD) survey, a seeing-limited optical 3D spectroscopy study of the central regions of all 170 galaxies in the Southern hemisphere with $B < 12.0$ and $|b| > 15\degr$. Most of the observations were taken with the Integral Field Unit of the Gemini Multi-Object Spectrograph, at the Gemini South telescope, but some are also being taken with the Southern Astrophysical Research Telescope (SOAR) Integral Field Spectrograph. The \divingTD survey was designed for the study of nuclear emission-line properties, circumnuclear (within scales of hundreds of pc) emission-line properties, stellar and gas kinematics and stellar archaeology. The data have a combination of high spatial and spectral resolution not matched by previous surveys and will result in significant contributions for studies related to, for example, the statistics of low-luminosity active galactic nuclei, the ionization mechanisms in Low-Ionization Nuclear Emission-Line Regions, the nature of transition objects, among other topics.

\end{abstract}

\begin{keywords}
galaxies: nuclei -- galaxies: active -- galaxies: Seyfert -- techniques: imaging spectroscopy 
\end{keywords}



\section{Introduction}\label{sec1}

Galaxies have been known as entities containing hundreds of billions of stars -- ``islands in the universe'' -- for nearly a century. Their nuclei certainly preserve important information about their origin and evolution. For these reasons it is important to study them, both at the individual level and on a statistical basis. Many galactic nuclei present emission lines associated with the accretion of matter onto a central supermassive black hole (SMBH), with a mass in the range of $10^6$--$10^{10}$ M$_{\sun}$. They are the so-called active galactic nuclei (AGNs). The luminosity function of AGNs is such that they can be studied at large distances. Curiously the most abundant objects are low luminosity AGNs (LLAGNs), which are not so well understood, despite their predominance in galaxies in the local Universe. The LLAGNs (as all AGNs) can be classified depending on the presence (type 1) or absence (type 2) of broad permitted emission lines. The detection of a broad component is considered a conclusive proof that the object contains an AGN and, therefore, a SMBH (see \citealt{net13} for a detailed discussion).

Most of the massive galaxies in the local Universe host an active nucleus \citep{ho08}, the majority of them presenting low-ionization emission lines and hence classified as Low-Ionization Nuclear Emission-Line Regions (LINERs; \citealt{hec80}). The nuclear emission in LINERs has been proposed to be similar to that of Seyfert galaxies, but in an environment with lower ionization parameter \citep{fer83,hal83}; this idea was confirmed by the discovery of broad H$\alpha$ emission in a significant fraction of LINERs \citep{ho97} as well as the detection of optical non-thermal continuum \citep{mao95}, high-ionization forbidden lines \citep{dud09} and compact central X-ray sources \citep{gon09}. Such characteristics are usually associated with AGNs. However it was also found that low-ionization emission can be quite extended in early-type galaxies (ETGs, \citealt{phi86, sar10}), far beyond the ionization produced by a low-luminosity central source. The source of this ionization was proposed to be a population of hot low-mass evolved stars (HOLMES\footnote{This term has been proposed by \citet{FloresFajardo.etal.2011a}; sometimes they are referred to as hot post-asymptotic giant branch stars, but it can also include nuclei of planetary nebulae and white dwarfs.}, \citealt{bin94}). In the last two decades significant evidence of nuclear activity has been detected in LINER galaxies (e.g. \citealt{nag05,ho08,mas11,hec14,caz18}); however, the evidence of extended HOLMES-ionized emission has also grown (e.g.  \citealt{sta08,era10,sar10,cid11,yan12,sin13,bel16}). 
Approximately two thirds of E--Sb galaxies exhibit local weak nuclear activity incompatible with normal stellar processes; in contrast, only about $15$ per cent of the Sc--Sm galaxies are known to have AGN activity \citep{ho08}. Late type galaxies are generally of low mass, gas rich, with strong star formation, and are frequently characterized as having a central stellar cluster \citep{wal06,bek10}. The nature of these clusters is still poorly understood and they are even considered as failed black holes \citep{neu12}. 

It is now well established that there are correlations between the masses of central SMBHs and certain properties of the host galaxies, such as the bulge stellar velocity dispersion (the so-called $M$--$\sigma$ relation; \citealt{fer00,geb00,gul09}), which has generated great interest in studying the connections between SMBH growth and galaxy formation/evolution. As a direct manifestation of accretion and growth, black holes have been considered as essential components of structure formation \citep{spr05,hop06,gra04}. 

An effective way of studying galaxies and their nuclei is by performing surveys of large samples. With such surveys, new and interesting objects can be found and, if the samples are selected by rigorous criteria, statistical properties can be derived. One of the most popular surveys of galactic nuclei was the PALOMAR survey \citep{fil85,ho08}. This survey was based on single spectra taken with a 2 arcsec $\times$ 4 arcsec slit on the Palomar 5 m telescope for every galaxy brighter than $B = 12.5$ in the Revised Shapley-Ames Catalog of Bright Galaxies (RSA, \citealt{san81}). A total of 486 galaxies was included in this survey. 

Important and influential as it was (and still is), the PALOMAR survey offers no information on the spatial distribution of the light-emitting/-absorbing sources in the central region of galaxies. That requires 3D spectroscopy with a high spatial resolution. To overcome this issue, we present the Deep IFS View of Nuclei of Galaxies (\divingTD) survey, which uses seeing-limited optical 3D spectroscopy data to study the nuclear and circumnuclear regions (scales of $\sim$ 100 pc) of all 170 galaxies in the Southern hemisphere with $B < 12.0$ and $|b| > 15\degr$. The combination of a statistical complete survey with such spatial scales is currently an unique opportunity to redraw the statistics of the ionization sources for LLAGNs.

The paper is organized as follows. In Section~\ref{sec2}, we describe in detail the \divingTD survey and its main objectives. In Section~\ref{sec3}, we present the methodologies used for the treatment and analysis of the observational data obtained for the survey. In Section~\ref{sec4}, we show some of the early results obtained so far. Finally, we present a summary of all topics in Section~\ref{sec5}.

\begin{small}
\onecolumn
\begin{longtable}{cccccccc}
\caption{Parameters of the galaxies in the \divingTD sample.\label{tbl1}}  \\
\hline
Object     & Type         & $B$   & $M_{B}$ & Distance (Mpc)                    & Obs. Date  & Grating    & Exp. Time (s)     \\ \hline
\multicolumn{8}{c}{GMOS-North}                                                                                            \\ \hline
NGC   4030 & SA(s)bc      & 11.07 & -21.38  & 29.9 $^a$ & 09/04/2013 & R831-G5302 & 3$\times$960  \\
NGC 4697   & E6           & 10.11 & -20.4   & 12.4 $^a$                              & 09/04/2014 & R831-G5302 & 3$\times$960  \\
NGC 4775   & SA(s)d       & 11.74 & -18.45  & 12 $^b$                               & 24/06/2016 & R831-G5302 & 3$\times$862  \\
NGC 4781   & SB(rs)d      & 11.69 & -18.85  & 12.8 $^c$                              & 27/06/2016 & R831-G5302 & 3$\times$862  \\
NGC 4981   & SAB(r)bc     & 11.83 & -19.54  & 19.4 $^c$                           & 24/06/2016 & R831-G5302 & 3$\times$862  \\
NGC 5792   & SB(rs)b      & 11.72 & -20.23  & 25 $^c$                                 & 23/06/2016 & R831-G5302 & 3$\times$862  \\
NGC 7184   & SB(r)c       & 11.67 & -21.07  & 36.1 $^c$                               & 20/06/2016 & R831-G5302 & 3$\times$862  \\
NGC 7314   & SAB(rs)bc    & 11.65 & -19.66  & 16 $^c$                                 & 06/07/2016 & R831-G5302 & 3$\times$862  \\ \hline
\multicolumn{8}{c}{GMOS-South}                                                                                            \\ \hline
NGC 134    & SAB(s)bc     & 10.96 & -20.41  & 18.3 $^c$                               & 24/06/2015 & R831-G5322 & 3$\times$930  \\
NGC 157    & SAB(rs)bc    & 11.04 & -19.64  & 12.9 $^d$                              & 23/12/2014 & R831-G5322 & 3$\times$930  \\
NGC 247    & SAB(s)d      & 9.51   & -17.88  & 3.34 $^e$                              & 26/06/2015 & R831-G5322 & 3$\times$930  \\
NGC 253    & SAB(s)c      & 8.13  & -19     & 3.05 $^e$                              & 07/10/2013 & R831-G5322 & 3$\times$910  \\
NGC 289    & SB(rs)bc     & 11.81 & -20.07  & 21.6 $^c$                              & 26/09/2016 & R831-G5322 & 3$\times$866  \\
NGC 300    & SA(s)d       & 8.70  & -17.38  & 2.17 $^c$                              & 26/08/2013 & B600-G5323 & 1800          \\
NGC 584    & E4           & 11.20  & -21.83  & 20 $^a$                               & 24/12/2013 & B600-G5323 & 1800          \\
NGC 596    & cD           & 11.88 & -20.28  & 21.7 $^a$                              & 24/12/2016 & B600-G5323 & 3$\times$565  \\
NGC 613    & SB(rs)bc     & 10.75  & -21.02  & 21.4 $^f$                              & 25/01/2015 & R831-G5322 & 3$\times$930  \\
NGC 720    & E5           & 11.15 & -21.63  & 30.7 $^g$                              & 11/08/2013 & B600-G5323 & 1800          \\
NGC 908    & SA(s)c       & 10.87 & -20.5   & 18.5 $^c$                              & 01/12/2014 & R831-G5322 & 3$\times$930  \\
NGC 936    & SB0(rs)      & 11.19 & -20.71  & 19.9 $^h$                              & 24/11/2014 & B600-G5323 & 3$\times$630  \\
NGC 1052   & E4           & 11.53 & -20.19  & 19.2 $^a$                              & 30/09/2013 & B600-G5323 & 1800          \\
NGC 1068   & (R)SA(rs)b   & 9.55  & -20.35  & 10.1 $^f$                              & 13/11/2010 & B600-G5323 & 6$\times$830  \\
NGC 1097   & SB(s)b       & 10.16 & -20.71  & 15.4 $^c$                             & 10/08/2016 & B600-G5323 & 3$\times$565  \\
NGC 1187   & SB(r)c       & 10.93 & -20.26  & 21.4 $^c$                             & 23/12/2014 & R831-G5322 & 3$\times$930  \\
NGC 1201   & SA0(r)       & 11.56 & -19.97  & 20.2 $^i$                             & 02/10/2016 & B600-G5323 & 3$\times$565  \\
NGC 1255   & SAB(rs)bc    & 11.60 & -19.19  & 14.1 $^c$                               & 29/12/2016 & R831-G5322 & 3$\times$866  \\
NGC 1291   & (R)SB0/a(s)  & 9.42  & -19.99  & 8.6 $^j$                              & 27/11/2014 & B600-G5323 & 3$\times$630  \\
NGC 1300   & SB(rs)bc     & 11.10 & -19.85  & 14.5 $^f$                               & 11/09/2013 & B600-G5323 & 1800          \\
NGC 1302   & (R)SB0/a(r)  & 11.38 & -18.48  & 8.36 $^b$                               & 07/01/2016 & B600-G5323 & 3$\times$565  \\
NGC 1313   & SB(s)d       & 9.37  & -18.33  & 3.7 $^j$                               & 04/12/2012 & B600-G5323 & 3$\times$589  \\
NGC 1316   & SAB0(s)      & 9.60  & -22.64  & 20.8 $^k$                               & 07/10/2013 & B600-G5323 & 1800          \\
NGC 1326   & (R)SB0(r)    & 11.34 & -20.29  & 18.9 $^l$                               & 28/09/2016 & B600-G5323 & 3$\times$565  \\
NGC 1332   & S0(s)        & 11.29 & -20.68  & 22.9 $^i$                               & 07/10/2013 & B600-G5323 & 1800          \\
NGC 1344   & E5           & 11.28 & -20.53  & 21 $^m$                               & 01/10/2016 & B600-G5323 & 3$\times$565  \\
NGC 1365   & SB(s)b       & 10.21 & -20.66  & 13.6 $^c$                               & 01/12/2014 & B600-G5323 & 3$\times$630  \\
NGC 1380   & SA0          & 11.10 & -21     & 20.6 $^m$                             & 07/08/2008 & B600-G5323 & 1800          \\
NGC 1387   & SAB0(s)      & 11.83 & -19.93  & 19.1 $^a$                             & 14/01/2016 & B600-G5323 & 3$\times$565  \\
NGC 1395   & E2           & 11.18 & -21.57  & 24.3 $^h$                              & 26/08/2013 & B600-G5323 & 1800          \\
NGC 1398   & (R')SB(r)ab  & 10.60 & -22.16  & 28.6 $^c$                             & 13/09/2015 & B600-G5323 & 3$\times$565  \\
NGC 1399   & E1           & 10.79    & -21.67  & 21.1 $^m$                              & 04/08/2008 & B600-G5323 & 1800          \\
NGC 1404   & E1           & 11.06 & -20.81  & 19  $^g$                                & 05/08/2008 & B600-G5323 & 1800          \\
NGC 1407   & E0           & 10.93 & -21.87  & 25.1 $^n$                              & 26/08/2013 & B600-G5323 & 1800          \\
NGC 1411   & SA0(r)       & 11.70 & -19.92  & 19.1 $^a$                              & 15/01/2016 & B600-G5323 & 3$\times$565  \\
NGC 1427   & cD           & 11.94 & -20.63  & 25.8 $^g$                              & 28/09/2016 & B600-G5323 & 3$\times$565  \\
NGC 1433   & (R')SB(r)ab  & 10.68 & -19.26  & 8.32 $^e$                              & 20/01/2016 & B600-G5323 & 3$\times$565  \\
NGC 1515   & SAB(s)bc     & 11.93 & -19.61  & 17.9 $^c$                              & 11/11/2016 & R831-G5322 & 3$\times$866  \\
NGC 1527   & SAB0(r)      & 11.70 & -19.74  & 18.5 $^h$                              & 16/12/2015 & B600-G5323 & 3$\times$565  \\
NGC 1533   & SB0          & 11.71 & -20.36  & 24.1 $^g$                             & 15/01/2016 & B600-G5323 & 3$\times$565  \\
NGC 1537   & SAB0         & 11.62 & -20.39  & 23.1 $^a$                              & 15/01/2016 & B600-G5323 & 3$\times$565  \\
NGC 1543   & (R)SB0(s)    & 11.49 & -20.81  & 18.4 $^g$                              & 20/01/2016 & B600-G5323 & 3$\times$565  \\
NGC 1549   & E0-1         & 10.76  & -20.51  & 16.3 $^g$                              & 28/08/2013 & B600-G5323 & 1800          \\
NGC 1553   & SA0(r)       & 10.42  & -19.99  & 9.51 $^g$                              & 24/11/2014 & B600-G5323 & 3$\times$630  \\
NGC 1559   & SB(s)cd      & 10.97   & -19.43  & 11.6 $^c$                              & 12/01/2014 & R831-G5322 & 3$\times$1000 \\
NGC 1566   & SAB(s)bc     & 10.21 & -21.02  & 18 $^o$                                & 10/10/2013 & R831-G5322 & 3$\times$910  \\
NGC 1574   & SA0(s)       & 11.19 & -20.69  & 20.1 $^h$                              & 28/08/2013 & B600-G5323 & 1800          \\
NGC 1617   & SB(s)a       & 11.37 & -19.48  & 13.4 $^j$                              & 01/10/2016 & B600-G5323 & 3$\times$565  \\
NGC 1672   & SB(s)b       & 11.03 & -19.98  & 11.4 $^p$                              & 15/01/2016 & B600-G5323 & 3$\times$565  \\
NGC 1700   & E4           & 11.96 & -20.65  & 33.1 $^g$                              & 04/11/2013 & B600-G5323 & 1800          \\
NGC 1792   & SA(rs)bc     & 10.85 & -19.37  & 10.8 $^c$                              & 20/12/2014 & R831-G5322 & 3$\times$930  \\
NGC 1808   & (R)SAB(s)a   & 10.70 & -19.35  & 9.51 $^c$                              & 15/12/2015 & B600-G5323 & 3$\times$565  \\
NGC 1947   & S0           & 11.86 & -20.37  & 19.6 $^q$                              & 15/12/2015 & B600-G5323 & 3$\times$565  \\
NGC 2207   & SAB(rs)bc    & 11.35 & -21.54  & 36.1 $^q$                              & 10/11/2016 & R831-G5322 & 3$\times$866  \\
\hline
Object     & Type         & $B$   & $M_{B}$ & Distance (Mpc)                    & Obs. Date  & Grating    & Exp. Time (s)     \\ \hline
NGC 2217   & (R)SB0(rs)   & 11.59 & -20.91  & 25.6 $^g$                             & 11/10/2013 & B600-G5323 & 1800          \\
NGC 2442   & SAB(s)bc     & 11.16 & -20.99  & 20.1 $^r$                              & 23/02/2014 & R831-G5322 & 3$\times$815  \\
NGC 2784   & SA0(s)       & 11.21 & -19.48  & 9.59 $^a$                              & 16/05/2013 & B600-G5323 & 1800          \\
NGC 2835   & SB(rs)c      & 10.95 & -18.82  & 8.75 $^a$                              & 12/05/2015 & R831-G5322 & 3$\times$865  \\
NGC 2974   & E4           & 11.78 & -20.36  & 22.1 $^a$                              & 28/06/2013 & B600-G5323 & 1800          \\
NGC 2997   & SAB(rs)c     & 10.32 & -20.94  & 13.1 $^f$                              & 27/11/2013 & R831-G5322 & 3$\times$930  \\
NGC 3115   & S0           & 9.98  & -20.17  & 9.68 $^i$                              & 30/06/2013 & B600-G5323 & 1800          \\
NGC 3223   & SA(s)b       & 11.88 & -21.54  & 37.5 $^c$                              & 07/02/2017 & B600-G5323 & 3$\times$560  \\
NGC 3511   & SA(s)c       & 11.56 & -19.2   & 12.6 $^c$                              & 11/05/2015 & R831-G5322 & 3$\times$865  \\
NGC 3513   & SB(rs)c      & 11.99 & -17.56  & 7.8 $^f$                               & 17/02/2018 & R831-G5322 & 3$\times$885  \\
NGC 3557   & E3           & 11.46 & -21.07  & 26.8 $^g$                              & 16/06/2013 & B600-G5323 & 1800          \\
NGC 3585   & E6           & 10.93 & -20.27  & 13.1 $^g$                              & 16/05/2013 & B600-G5323 & 1800          \\
NGC 3621   & SA(s)d       & 10.03 & -19.29  & 6.4 $^c$                               & 22/02/2014 & R831-G5322 & 3$\times$815  \\
NGC 3672   & SA(s)c       & 11.66 & -20.2   & 25.1 $^c$                              & 17/03/2017 & R831-G5322 & 3$\times$870  \\
NGC 3887   & SB(r)bc      & 11.60  & -19.92  & 18.2 $^f$                              & 17/03/2017 & R831-G5322 & 3$\times$870  \\
NGC 3904   & E2-3         & 11.95 & -20.81  & 28.3 $^a$                              & 11/05/2013 & B600-G5323 & 1800          \\
NGC 3923   & E4-5         & 10.91 & -20.86  & 16 $^g$                                & 14/05/2013 & B600-G5323 & 1800          \\
NGC 3962   & E1           & 11.66 & -21.49  & 36.3 $^a$                              & 29/06/2013 & B600-G5323 & 1800          \\
NGC 4105   & E3           & 11.88 & -20.79  & 26.7 $^a$                              & 01/07/2013 & B600-G5323 & 1800          \\
NGC 4487   & SAB(rs)cd    & 11.66 & -18.62  & 12.3 $^c$                              & 09/03/2017 & R831-G5322 & 3$\times$870  \\
NGC 4504   & SA(s)cd      & 11.92 & -19.24  & 17.5 $^c$                            & 23/02/2018 & R831-G5322 & 3$\times$885  \\
NGC 4546   & SB0(s)       & 11.30 & -19.46  & 14 $^a$                                & 17/02/2008 & B600-G5323 & 1200          \\
NGC 4593   & (R)SB(rs)b   & 11.72 & -20.92  & 37.2 $^q$                             & 02/03/2017 & B600-G5323 & 3$\times$560  \\
NGC 4594   & SA(s)a       & 9.28  & -22.72  & 21.7 $^s$                              & 03/02/2011 & B600-G5323 & 3$\times$595  \\
NGC 4666   & SABc         & 11.56 & -19.53  & 14 $^c$                                & 11/05/2015 & R831-G5322 & 3$\times$882  \\
NGC 4691   & (R)SB0/a(s)  & 11.70 & -20.11  & 20.6 $^b$                             & 13/02/2015 & B600-G5323 & 3$\times$582  \\
NGC 4696   & cD1          & 11.59  & -22.2   & 39.8 $^n$                             & 04/05/2013 & B600-G5323 & 1800          \\
NGC 4699   & SAB(rs)b     & 10.44 & -21.24  & 19.5 $^c$                              & 16/05/2013 & B600-G5323 & 1800          \\
NGC 4731   & SB(s)cd      & 11.55  & -18.92  & 13.2 $^c$                              & 11/05/2015 & R831-G5322 & 3$\times$865  \\
NGC 4753   & I0           & 10.85 & -21.49  & 24.1 $^a$                              & 11/05/2015 & B600-G5323 & 3$\times$565  \\
NGC 4818   & SAB(rs)ab    & 11.89 & -18.22  & 11.3 $^c$                             & 24/02/2017 & B600-G5323 & 3$\times$560  \\
NGC 4902   & SB(r)b       & 11.90 & -20.82  & 32.4 $^b$                              & 03/02/2017 & B600-G5323 & 3$\times$560  \\
NGC 4939   & SA(s)bc      & 11.56  & -21.39  & 42.5 $^c$                              & 14/02/2015 & R831-G5322 & 3$\times$865  \\
NGC 4941   & (R)SAB(r)ab  & 11.90 & -19.72  & 21.2 $^t$                              & 09/02/2017 & B600-G5323 & 3$\times$560  \\
NGC 4958   & SB0(r)       & 11.48 & -20.27  & 21.4 $^q$                              & 12/05/2015 & B600-G5323 & 3$\times$582  \\
NGC 4984   & (R)SAB0(rs)  & 11.71 & -20.1   & 21.3 $^j$                              & 08/08/2016 & B600-G5323 & 570           \\
NGC 4995   & SAB(rs)b     & 11.90  & -20.1   & 23.4 $^c$                              & 02/03/2017 & B600-G5323 & 3$\times$560  \\
NGC 5018   & E3           & 11.71 & -21.48  & 37.5 $^u$                              & 08/06/2013 & B600-G5323 & 1800          \\
NGC 5044   & E0           & 11.92 & -21.15  & 32.2 $^a$                              & 04/05/2013 & B600-G5323 & 1800          \\
NGC 5054   & SA(s)bc      & 11.51 & -20.04  & 18.2 $^c$                              & 29/04/2015 & R831-G5322 & 3$\times$865  \\
NGC 5061   & E0           & 11.35 & -20.07  & 15.2 $^g$                              & 02/03/2015 & B600-G5323 & 3$\times$565  \\
NGC 5068   & SAB(rs)cd    & 10.53 & -18.92  & 6.7 $^j$                               & 29/04/2015 & R831-G5322 & 3$\times$865  \\
NGC 5101   & (R)SB0/a(rs) & 11.58 & -19.73  & 14 $^g$                               & 11/05/2013 & B600-G5323 & 1800          \\
NGC 5102   & SA0          & 10.64 & -17.29  & 4 $^i$                                 & 02/03/2015 & B600-G5323 & 3$\times$565  \\
NGC 5128   & S0           & 7.89   & -19.75  & 3.53 $^v$                             & 29/04/2015 & B600-G5323 & 3$\times$565  \\
NGC 5161   & SA(s)c       & 11.98 & -19.73  & 21.9 $^c$                              & 17/02/2018 & R831-G5322 & 3$\times$885  \\
NGC 5170   & SA(s)c       & 11.88 & -20.43  & 27.5 $^c$                              & 09/03/2017 & R831-G5322 & 3$\times$870  \\
NGC 5236   & SAB(s)c      & 8.51  & -20.95  & 4.79 $^w$                              & 23/02/2014 & R831-G5322 & 3$\times$815  \\
NGC 5247   & SA(s)bc      & 11.10 & -21.06  & 22.2 $^j$                              & 29/04/2015 & R831-G5322 & 3$\times$865  \\
NGC 5334   & SB(rs)c      & 11.90 & -20.46  & 32.6 $^x$                              & 10/03/2017 & R831-G5322 & 3$\times$875  \\
NGC 5556   & SAB(rs)d     & 11.88 & -18.95  & 16.1 $^a$                             & 08/03/2017 & R831-G5322 & 3$\times$870  \\
NGC 5584   & SAB(rs)cd    & 11.95 & -18.64  & 17.9 $^c$                             & 19/02/2018 & R831-G5322 & 3$\times$885  \\
NGC 5643   & SAB(rs)c     & 10.89 & -20.7   & 16.9 $^j$                             & 22/02/2014 & R831-G5322 & 3$\times$815  \\
NGC 6118   & SA(s)cd      & 11.91 & -19.71  & 20.7 $^c$                              & 27/02/2018 & R831-G5322 & 3$\times$885  \\
NGC 6684   & (R')SB0(s)   & 11.34 & -19.55  & 13.6 $^a$                              & 30/04/2015 & B600-G5323 & 3$\times$565  \\
NGC 6744   & SAB(r)bc     & 9.24  & -19.71  & 7.66 $^c$                              & 08/05/2014 & R831-G5322 & 3$\times$815  \\
NGC 6868   & E2           & 11.83 & -21.47  & 34 $^g$                                & 04/05/2013 & B600-G5323 & 1800          \\
NGC 7049   & SA0(s)       & 11.64 & -21.46  & 29.9 $^a$                             & 11/05/2013 & B600-G5323 & 1800          \\
NGC 7090   & SBc          & 11.10 & -17.58  & 6.22 $^c$                              & 26/07/2015 & R831-G5322 & 3$\times$884  \\
NGC 7144   & E0           & 11.79 & -20.25  & 24.4 $^a$                             & 24/07/2015 & B600-G5323 & 3$\times$565  \\
NGC 7205   & SA(s)bc      & 11.57 & -19.85  & 18.3 $^c$                             & 09/08/2016 & R831-G5322 & 3$\times$861  \\
NGC 7213   & SA(s)a       & 11.18 & -20.65  & 22 $^j$                                & 12/05/2015 & B600-G5323 & 3$\times$565  \\
NGC 7410   & SB(s)a       & 11.30 & -21.91  & 44.1 $^c$                              & 12/05/2015 & B600-G5323 & 3$\times$565  \\
NGC 7424   & SAB(rs)cd    & 10.99 & -19.89  & 11.5 $^j$                              & 23/09/2013 & R831-G5322 & 3$\times$808  \\
NGC 7496   & SB(s)b       & 11.78 & -17.39  & 7.4 $^s$                               & 31/05/2017 & B600-G5323 & 3$\times$560  \\
NGC 7507   & E0           & 11.43 & -21.52  & 24.6 $^a$                              & 10/08/2013 & B600-G5323 & 1800          \\
\hline
Object     & Type         & $B$   & $M_{B}$ & Distance (Mpc)                    & Obs. Date  & Grating    & Exp. Time (s)     \\ \hline
NGC 7552   & (R')SB(s)ab  & 11.40 & -19.72  & 14.8 $^y$                              & 04/08/2017 & B600-G5323 & 3$\times$560  \\
NGC 7582   & (R')SB(s)ab  & 11.46 & -20.48  & 22.5 $^a$                              & 17/07/2004 & B600-G5323 & 3$\times$720  \\
NGC 7606   & SA(s)b       & 11.55 & -21.26  & 37.3 $^c$                             & 16/06/2016 & B600-G5323 & 3$\times$570  \\
NGC 7713   & SB(r)d       & 11.65 & -18.33  & 9.25 $^c$                             & 13/07/2018 & R831-G5322 & 3$\times$885  \\
NGC 7793   & SA(s)d       & 9.65   & -18.23  & 3.73 $^c$                              & 10/10/2016 & R831-G5322 & 3$\times$865  \\
IC 1459    & E3-4         & 10.96 & -21.76  & 28.7 $^a$                              & 03/08/2008 & B600-G5323 & 1800          \\
IC 4296    & E            & 11.58 & -22.43  & 50.8 $^{\alpha}$                            & 16/05/2013 & B600-G5323 & 1800          \\
IC 5201    & SB(rs)cd     & 11.54 & -18.44  & 10.5 $^c$                              & 12/05/2015 & R831-G5322 & 3$\times$865  \\
IC 5267    & SA0/a(s)     & 11.39 & -20.95  & 24.5 $^g$                              & 16/05/2015 & B600-G5323 & 3$\times$582  \\
IC 5273    & SB(rs)cd     & 11.90 & -19.12  & 15.5 $^c$                              & 01/07/2017 & R831-G5322 & 3$\times$870  \\
IC 5328    & E4           & 11.95 & -20.93  & 39.7 $^g$                              & 08/08/2016 & B600-G5323 & 3$\times$570  \\ \hline
\multicolumn{8}{c}{SIFS}                                                                                                  \\ \hline
NGC 150    & SB(rs)b      & 11.75    & -19.87  & 23.1 $^c$                              &     21/11/2019    & 700 l/mm   & 3$\times$1200              \\
NGC 210    & SAB(s)b      & 11.65 & -19.97  & 21.6 $^d$                              &     13/09/2018    & 700 l/mm   &  3$\times$1200      \\
NGC 578    & SAB(rs)c     & 11.48 & -19.62  & 18.1 $^c$                              &     26/11/2019    & 700 l/mm   &  3$\times$1200      \\
NGC 986    & SB(rs)ab     & 11.80 & -18.93  & 11.1 $^y$                              &     21/11/2019      & 700 l/mm   &  3$\times$1200     \\
NGC 1042   & SAB(rs)cd    & 11.49 & -16.44  & 4.2 $^{\beta}$                         &     27/11/2019    & 700 l/mm   &     3$\times$1200    \\
NGC 1087   & SAB(rs)c     & 11.56 & -19.17  & 12.9 $^c$                              &     27/11/2019    & 700 l/mm   &     3$\times$1200    \\
NGC 1232   & SAB(rs)c     & 10.50  & -20.2   & 14.5 $^f$                              &     23/11/2017    & 700 l/mm   &     3$\times$1200    \\
NGC 1350   & (R')SB(r)ab  & 11.40 & -20.44  & 19.7 $^c$                              &     13/09/2018    & 700 l/mm   &     3$\times$1200    \\
NGC 1371   & SAB(rs)a     & 11.50 & -21.13  & 30.3 $^c$                              &     21/11/2019    & 700 l/mm   &     3$\times$1200    \\
NGC 1385   & SB(s)cd      & 11.65 & -18.1   & 7.9 $^s$                               &     26/11/2019    & 700 l/mm 
&     3$\times$1200    \\
NGC 1421   & SAB(rs)bc    & 11.95 & -19.6   & 19.2 $^c$                              &     23/11/2019    & 700 l/mm 
&     3$\times$1200    \\
NGC 1425   & SA(s)b       & 11.60 & -20.6   & 22.5 $^c$                              &     26/11/2019    & 700 l/mm   &     3$\times$1200    \\
NGC 1448   & SAcd         & 11.30 & -19.67  & 17 $^c$                                &     23/11/2017    & 700 l/mm 
&     3$\times$1200    \\
NGC 1493   & SB(r)cd      & 11.82 & -18.45  & 11.3 $^j$                              &     23/11/2019    & 700 l/mm 
&     3$\times$1200    \\
NGC 1512   & SB(r)a       & 11.38 & -19.52  & 12.6 $^c$                              &     13/09/2018    & 700 l/mm   &     3$\times$1200    \\
NGC 1744   & SB(s)d       & 11.70  & -18.24  & 9.95 $^c$                              &     23/11/2019    & 700 l/mm 
&     3$\times$1200    \\
NGC 1964   & SAB(s)b      & 11.60  & -20.2   & 23.7 $^c$                              &     26/11/2019    & 700 l/mm   &     3$\times$1200    \\
NGC 2090   & SA(rs)c      & 11.85 & -19.25  & 13.4 $^c$                              &     23/11/2019    & 700 l/mm 
&     3$\times$1200    \\
NGC 4856   & SB0/a(s)     & 11.40  & -20.49  & 24 $^t$                                &     06/02/2017    & 700 l/mm   &     3$\times$1200    \\
NGC 5530   & SA(rs)bc     & 11.98 & -18.93  & 11.6 $^c$                              &     20/05/2018    & 700 l/mm 
&     3$\times$1200    \\
NGC 6753   & (R)SA(r)b    & 11.93 & -20.34  & 23.4 $^g$                              &     20/05/2018    & 700 l/mm 
&     3$\times$1200    \\
NGC 7083   & SA(s)bc      & 11.80  & -21.13  & 36 $^a$                                &     19/05/2018    & 700 l/mm 
&     3$\times$1200    \\
NGC 7723   & SB(r)b       & 11.85 & -20.55  & 29.2 $^c$                              &     13/09/2018    & 700 l/mm 
&     3$\times$1200    \\
NGC 7727   & SAB(s)a      & 11.55 & -20.26  & 19.7 $^b$                              &     13/09/2018    & 700 l/mm   &     3$\times$1200    \\
IC 5332    & SA(s)d       & 11.25 & -19.03  & 8.4 $^j$                               &     13/09/2018    & 700 l/mm 
&     3$\times$1200    \\ \hline
\multicolumn{8}{c}{Not Observed}                                                                                          \\ \hline
NGC 685    & SAB(r)c      & 11.97 & -19.02  & 15.2 $^j$                             &            &            &               \\
NGC 779    & SAB(r)b      & 11.86 & -19.69  & 20 $^c$                                &            &           &               \\
NGC 1084   & SA(s)c       & 11.25 & -20.23  & 17.4 $^c$                             &            &           &               \\
NGC 1249   & SB(s)cd      & 11.80 & -18.84  & 15.3 $^c$                            &            &            &               \\
NGC 1532   & SB(s)b       & 11.53  & -20.75  & 19.9 $^c$                             &            &             &               \\
NGC 1637   & SAB(rs)c     & 11.52 & -19.89  & 17.5 $^a$                             &            &             &               \\\hline
 
\multicolumn{8}{c}{Distances for the galaxies were obtained from: $^a$ -\citet{AAAA}, $^b$ -\citet{BBBB}, $^c$ - \citet{CCCC},}\\ 
\multicolumn{8}{c}{$^d$ - \citet{DDDD}, $^e$ - \citet{EEEE}, $^f$ - \citet{FFFF}, $^g$ - \citet{GGGG}}\\
\multicolumn{8}{c}{$^h$ - \citet{HHHH},  $^i$ - \citet{IIII}, $^j$ - \citet{JJJJ}, $^k$ - \citet{KKKK},}\\
\multicolumn{8}{c}{$^l$ - \citet{LLLL}, $^m$ - \citet{MMMM}, $^n$ - \citet{NNNN} }\\
\multicolumn{8}{c}{$^o$ - \citet{OOOO}, $^p$ - \citet{PPPP}, $^q$ - \citet{QQQQ}, $^r$ - \citet{RRRR}, $^s$ - \citet{SSSS},}\\
\multicolumn{8}{c}{$^t$ - \citet{TTTT}, $^u$ - \citet{UUUU}, $^v$ - \citet{VVVV}, $^w$ - \citet{WWWW},}\\
\multicolumn{8}{c}{$^x$ - \citet{XXXX}, $^y$ - \citet{YYYY}, $^{\alpha}$ - \citet{ALPHA}, $^{\beta}$ - \citet{BETA}. }
\end{longtable}

\end{small}

\twocolumn

\section{The \divingTD Survey}\label{sec2}

\begin{figure*}
    \centering
    \includegraphics[scale=0.5]{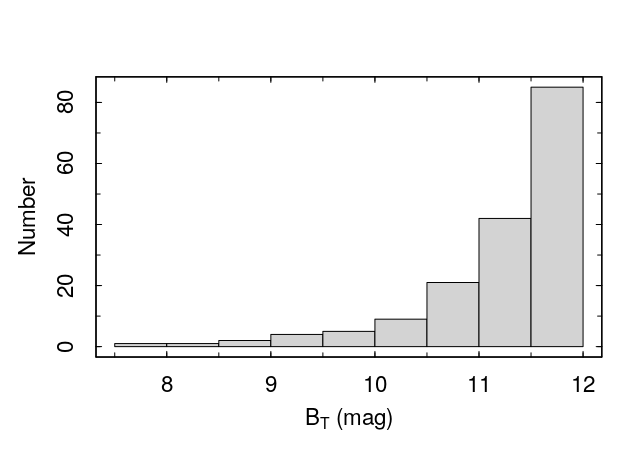}
    \includegraphics[scale=0.5]{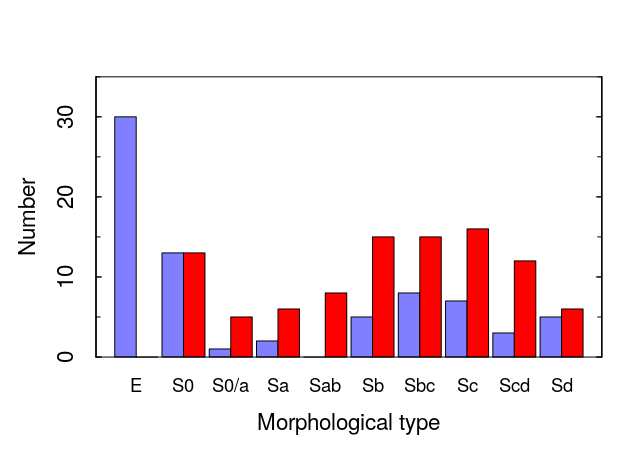}
    \includegraphics[scale=0.5]{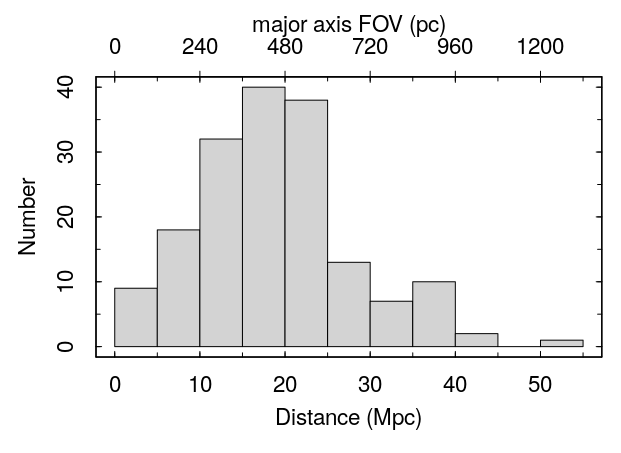}
    \includegraphics[scale=0.5]{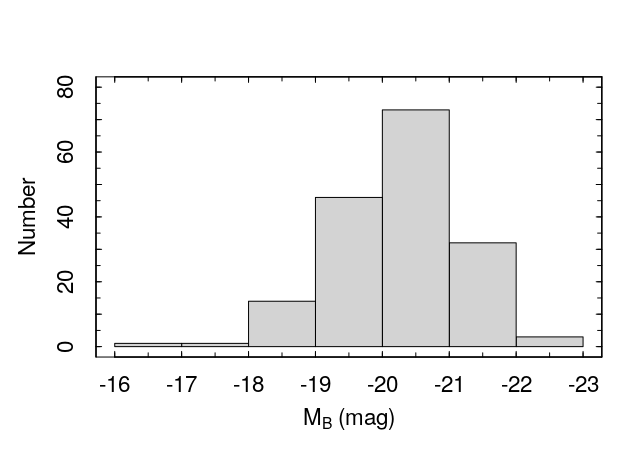}
    \caption{Basic properties of the \divingTD sample. Top left: distribution of apparent B magnitudes. Top right: distribution of morphological types. The purple bars correspond to unbarred galaxies and the red bars correspond to barred galaxies. Note that all elliptical galaxies of the sample do not have a bar and that all Sab galaxies have a bar, according to RC 3. Bottom left: distribution of distances, with a median value of 17.7 Mpc. Bottom right: distribution of absolute B magnitudes, with a median value of -20.06.}
    \label{fig01}
\end{figure*}

In order to build the \divingTD sample, we selected all galaxies from RSA in the Southern hemisphere ($\delta$ < 0) with $B < 12.0$ and $|b| > 15\degr$. From this selection, 11 Sm/Im objects were excluded, since it was not possible to clearly identify their nuclei in the 2MASS images. The final sample has 170 objects and is shown in Table~\ref{tbl1}. It is worth mentioning that RSA is statistically complete for $B < 12.0$ \citep{san81}. Fig. \ref{fig01} presents the distributions of the apparent and absolute $B$ magnitudes, distances and morphological types for the final \divingTD sample. The references for the distances of the galaxies are shown in Table \ref{tbl1}. The median distance of the objects in our sample is 19.0 Mpc. The apparent B magnitudes were taken from RSA. Whenever possible, the absolute $B$ magnitudes, corrected for Galactic extinction, were obtained from the Carnegie-Irvine Galaxy Survey (CGS; \citealt{ho11}), or else from Hyperleda \citep{pat03}. The median absolute $B$ magnitude for the sample is $-20.24$. The morphological types were taken from RC 3 \citep{dev91} and they are distributed as follows: 30 ellipticals (18 per cent), 32 S0+S0/a (19 per cent), 16 Sa+Sab (9 per cent), 43 Sb+Sbc (25 per cent), 23 Sc (14 per cent) and 26 Scd+Sd (15 per cent). Bars are present in almost two thirds of the disc galaxies (S0 plus spirals).

\begin{figure}
    \centering
    \includegraphics[scale=0.5]{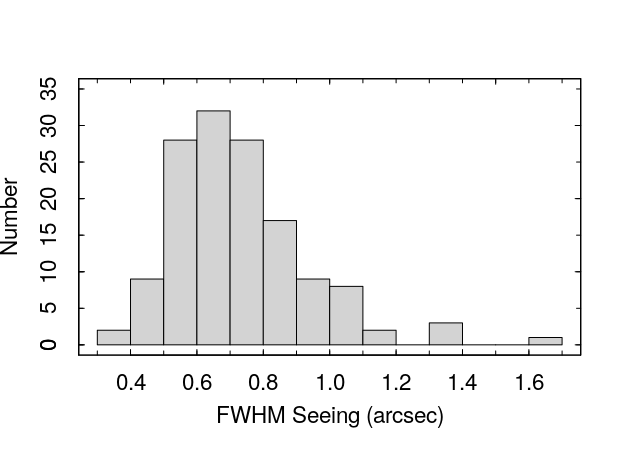}
    \caption{Distribution of the seeing values of all objects of the \divingTD sample that were observed with GMOS, with a median value of 0.70 arcsec.\label{fig02}}
    
\end{figure}

\begin{figure}
   \centering
   \includegraphics[scale=0.50]{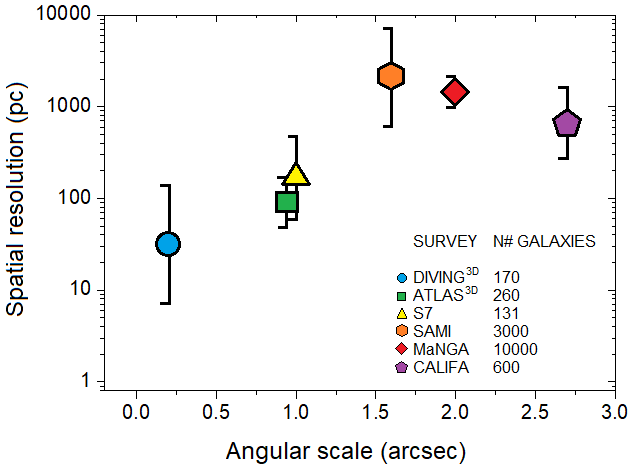} 
  \caption{Diagram showing the angular scale of the instruments (fibre or image slicer sizes) and the spatial resolutions between the nearest and the farthest galaxies of some IFU galaxy surveys (within the vertical brackets "]"). The surveys are identified in the bottom right part of the diagram (namely, the ATLAS$^{3D}$, S7, SAMI, MaNGA, CALIFA) and are followed by their estimated number of observed galaxies. Since the angular resolutions of the \divingTD survey are limited by the seeing of the observations, we opted to use the median seeing of 0.7 arcsec to calculate its spatial resolution. The spatial resolutions of the other surveys are limited by the angular scale of the instruments. One may see that the data cubes of the \divingTD survey reach the highest spatial resolutions so far.}\label{fig03}
\end{figure}

\begin{figure*}
   \centering
   \includegraphics[scale=0.72]{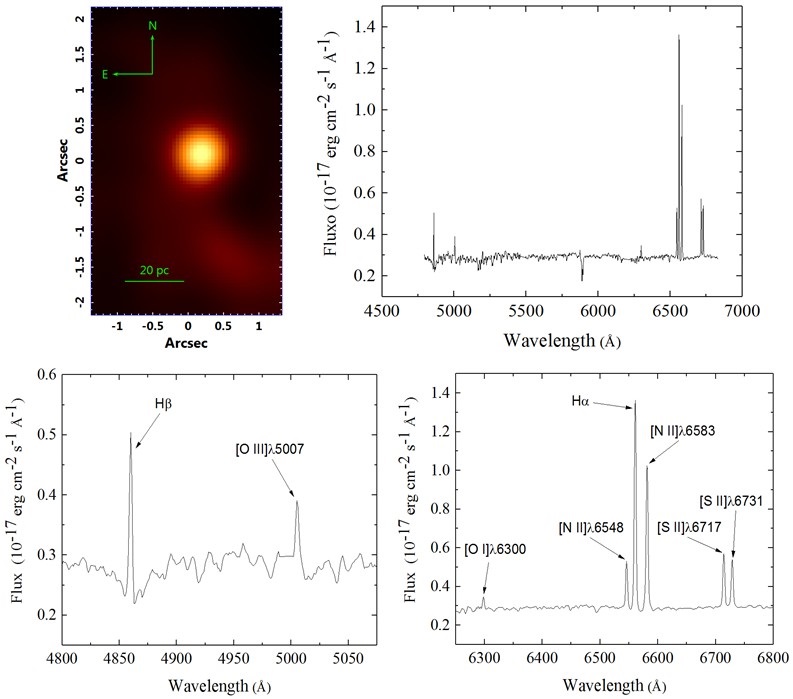} 
  \caption{Image of the GMOS/IFU data cube of the central region of the galaxy NGC 5236, collapsed along the spectral axis, and its average spectrum. Magnifications of the average spectrum are shown at the bottom.}\label{fig04}
\end{figure*}

\begin{figure*}
   \centering
   \includegraphics[scale=0.40]{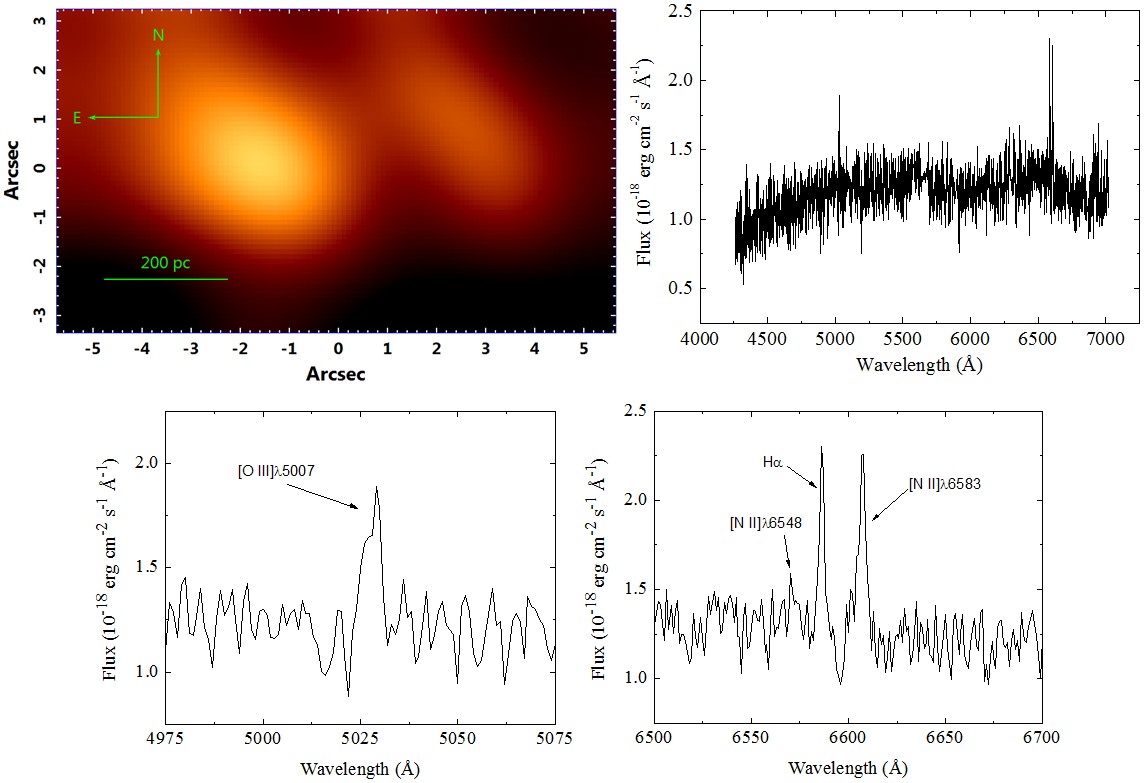} 
  \caption{Image of the SIFS data cube of the central region of the galaxy NGC 1448, collapsed along the spectral axis, and its average spectrum. Magnifications of the average spectrum are shown at the bottom.}\label{fig05}
\end{figure*}

\begin{figure*}
   \centering
   \includegraphics[scale=0.3]{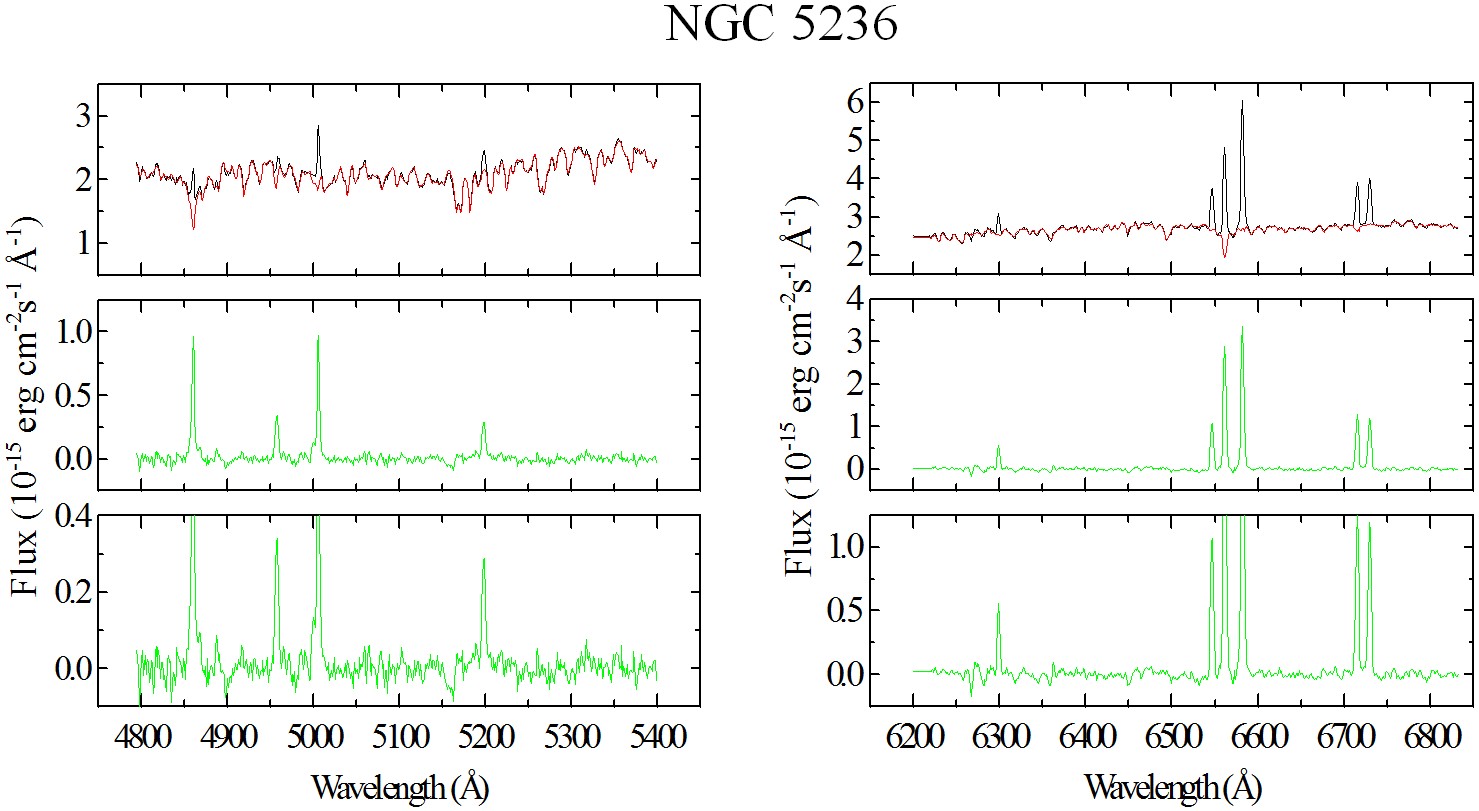} 
  \caption{Spectrum extracted from the nucleus of the galaxy NGC 5236. The observed spectrum is shown in black, the fit provided by the pPXF technique is shown in red and the fit residuals are shown in green.}\label{fig06}
\end{figure*}

\begin{figure*}
   \centering
   \includegraphics[scale=0.3]{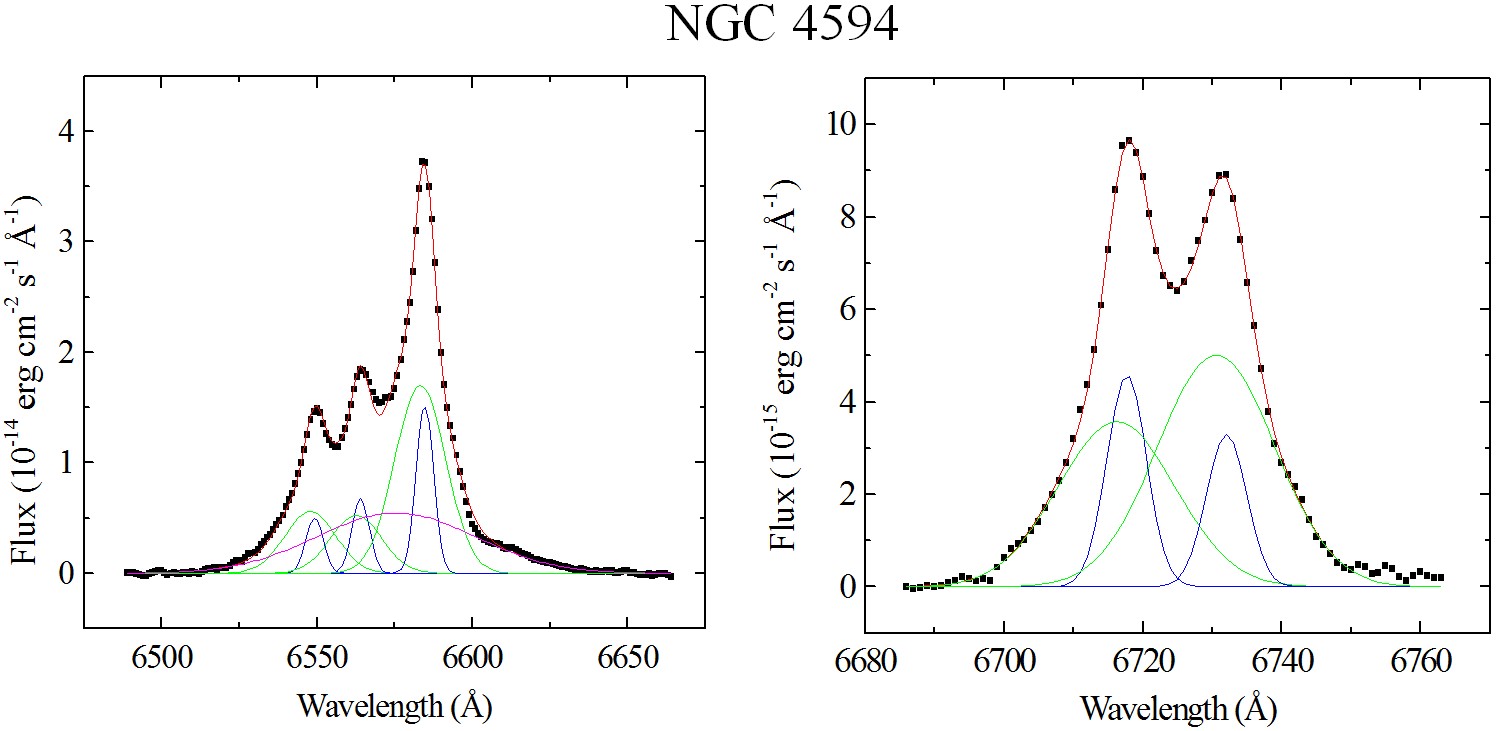} 
  \caption{Gaussian fits applied to the [S \textsc{ii}] and [N \textsc{ii}]+H$\alpha$ emission lines of the nuclear spectrum of NGC 4594. The green and blue curves correspond to the narrow components of the emission lines, the magenta curve represents the broad component of the H$\alpha$ emission line and the red curve corresponds to the final fit.}\label{fig07}
\end{figure*}

\begin{figure*}
   \centering
   \includegraphics[scale=0.32]{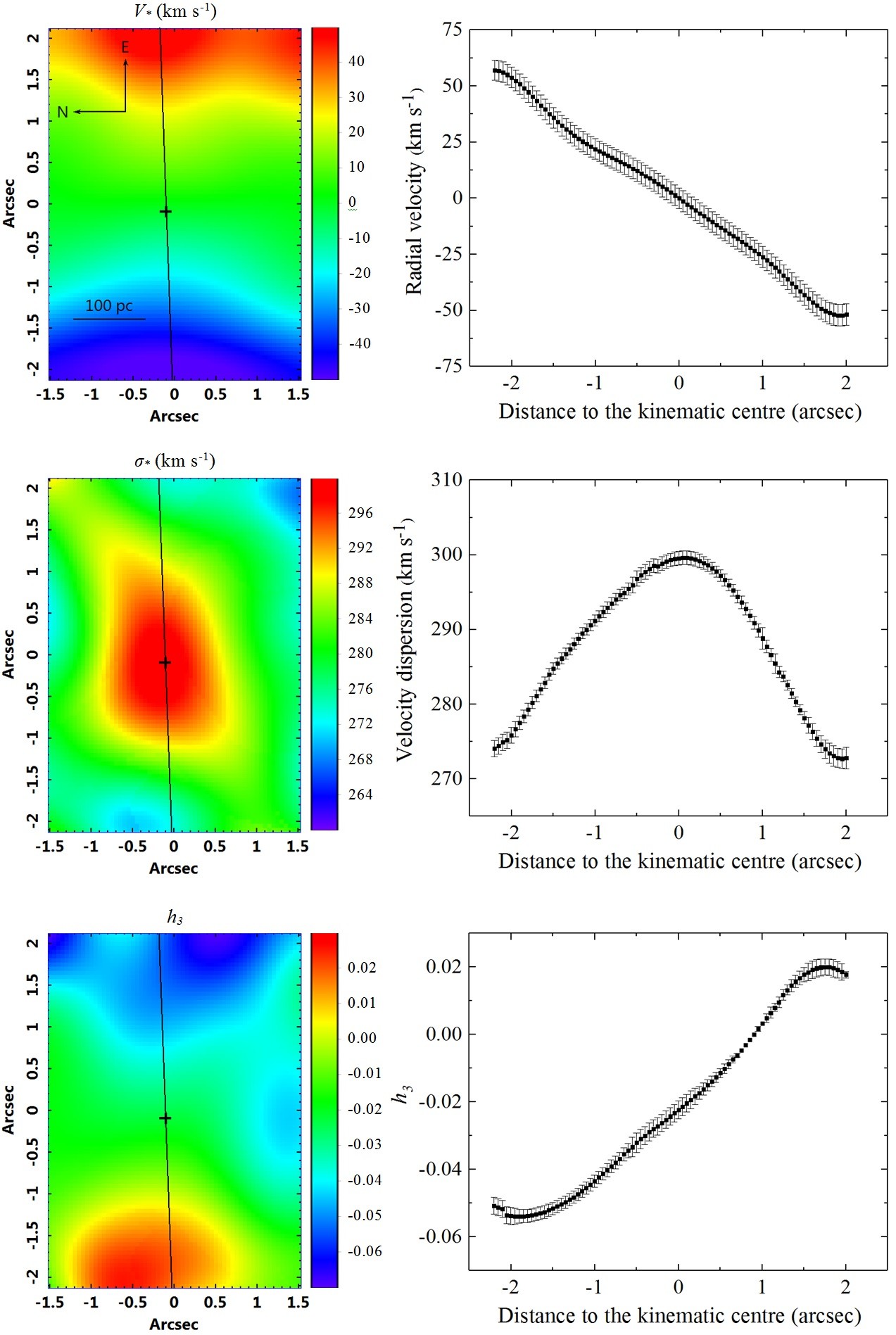} 
  \caption{Maps of $V_*$, $\sigma_*$ and of the Gauss-Hermite coefficient $h_3$ obtained with the pPXF technique applied to the GMOS/IFU data cube of the central region of NGC 1395. The curves of each map extracted from the axis corresponding to the line of nodes of the $V_*$ map, represented by the black line, are shown on the right. The kinematic centre of the $V_*$ map, corresponding to the black cross, was taken as the point, along the line of nodes, at which the velocity was equal to the average between the maximum and minimum velocities. Such average velocity was subtracted from the $V_*$ map, which shows then the velocity values relative to the kinematic centre.}\label{fig08}
\end{figure*}

\begin{figure*}
   \centering
   \includegraphics[scale=0.32]{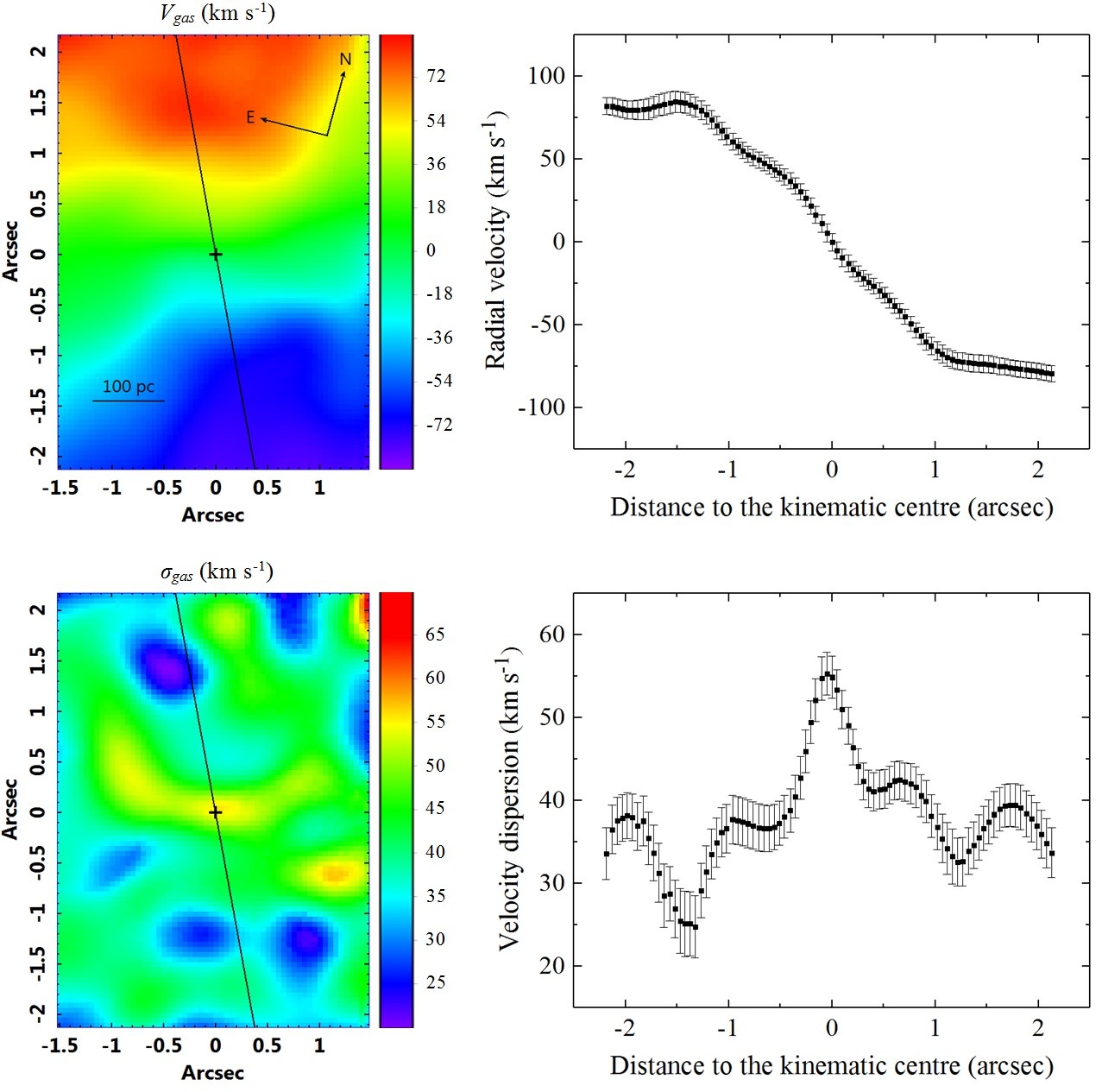} 
  \caption{Maps of $V_\mathrm{gas}$ and $\sigma_\mathrm{gas}$ obtained by fitting the [N \textsc{ii}]+H$\alpha$ emission lines, in the data cube of the central region of NGC 1395, with a sum of three Gaussian functions with the same width and radial velocity. The curves of each map, extracted from the axis corresponding to the line of nodes of the $V_\mathrm{gas}$ map, represented by the black line, are shown on the right. The kinematic centre of the $V_\mathrm{gas}$ map, corresponding to the black cross, was taken as the point, along the line of nodes, at which the velocity was equal to the average between the maximum and minimum velocities. Such average velocity was subtracted from the $V_\mathrm{gas}$ map, which shows then the velocity values relative to the kinematic centre.}\label{fig09}
\end{figure*}

A total of 131 galaxies were observed with the IFU of the Gemini Multi-Object Spectrograph (GMOS; \citealt{all02}) at the 8 m Gemini South telescope. Eight galaxies were observed using the same instrument, but at the Gemini-North telescope. The observations were taken in the one-slit mode, which resulted in a field of view (FOV), sampled by 500 fibers (with a diameter of 0.2 arcsec each), of 5 arcsec $\times$ 3.5 arcsec and in a sky FOV (at projected distance of 1 arcmin from the science FOV), sampled by 250 fibers, of 5 arcsec $\times$ 1.75 arcsec. 

The SOAR Integral Field Spectrograph (SIFS), at the 4 m SOAR telescope, was used to observe 25 galaxies from the sample. The fore-optics module of the instrument was set to produce a magnification of the image of 0.3 arcsec/fibre at the input of the IFU, resulting in a FOV of 7.8 arcsec $\times$ 15 arcsec. 

The GMOS/IFU and SIFS observations were all seeing limited. The median seeing values are 0.70 arcsec for the observations performed with GMOS/IFU and 1.04 arcsec for those obtained with SIFS. For the GMOS/IFU observations, we used field stars that are present in the acquisition image of the galaxy to estimate the seeing values. For SIFS, we also used acquisition images, but for the standard stars that were observed in the same night as the galaxies. We intend to observe the remaining six galaxies of the \divingTD sample with SIFS in the near future. It is worth emphasizing that, although this survey uses data obtained with GMOS/IFU and SIFS, such instruments provide very similar spectral coverages and spectral and spatial resolutions. Therefore, the use of these two instruments will not introduce significant biases in our analyses. Fig.~\ref{fig02} shows a histogram with the seeing values of all the observations taken with GMOS. More detail on the observations and data reduction of the \divingTD sample is presented in Section \ref{sec31}.

\subsection{The scientific rational}\label{sec21}

A distinguishing characteristic of the \divingTD survey, in comparison to other galaxy samples observed with IFUs, is the high spatial resolution. The idea is to have the highest possible spatial resolution and signal-to-noise ratio. Previous IFU surveys, such as CALIFA \citep{san12}, SAURON \citep{dez02}, ATLAS$^\mathrm{3D}$ \citep{cap11}, SAMI \citep{bry15}, S7 \citep{dop15,tho17} and MaNGA \citep{bun15} were able to spatially resolve phenomena that are all mixed up in Palomar and SDSS data. The \divingTD survey will, by contrast, map phenomena that are blurred in the data cubes of those surveys, i.e. physical processes in the nuclear and circumnuclear scales (100 pc) with a spatial resolution of $\sim$ 20 pc. No other IFU survey is aiming at this sweet-spot region of galaxies, where high stellar and gaseous densities, high metallicities, presence of (active or dormant) SMBHs and other extreme conditions drive a variety of phenomena seen nowhere else in galaxies: ionization cones, obscuring tori, inflows and outflows, nuclear clusters, inner gaseous and stellar disks, etc. In Fig.~\ref{fig03}, we present the spatial resolutions and the pixel scales of the \divingTD data in comparison to other IFU surveys. 

Recently, some works have shown the importance of using IFU data with high spatial resolution in order to study some of these phenomena. For instance, \citet{wyl17} used both GMOS and MaNGA data cubes to study ionized gas outflows in two AGN hosts. In particular, in the object with lower bolometric luminosity, the outflow was only revealed using the GMOS data cube. \citet{rif19} also used both GMOS and MaNGA data to characterize the prototype Red Geiser galaxy Akira. In this case, the GMOS data cube was important to detect a change of direction in the outflow from the inner part of the galaxy to kpc scales. In some other galaxies, a stellar kinematically decoupled core with scales of $\sim$ 100 pc was only detected when high-spatial resolution data cubes were used as a complement to large FOV IFU observations \citep{mcd06,ric16}. A systematic study of the circumnuclear regions using the \divingTD survey will be very important in order to characterize these structures from a statistical point of view.

The exquisite spatial resolution of the \divingTD survey will allow us to investigate the connection between AGNs and surrounding stellar populations to an unprecedented level of spatial detail, shedding new light on long-standing puzzles. For instance, while type 2 Seyferts show a clear tendency to host recent star-formation \citep{hec97,gon01,cid11b,kau03}, it is unclear whether this also happens in type 1 Seyferts. A natural extension of this topic is the co-evolution between the central SMBHs and their host galaxies, which is suggested by the existence of correlations between the masses of the SMBHs and parameters of the host galaxies. However, the existence and the way this co-evolution occurs are still uncertain. \citet{kor13}, for example, question whether low-mass galaxies follow a $M$--$\sigma$ relation and also argue that the co-evolution between the SMBHs and the host galaxies is probably more complex than suggested by the $M$--$\sigma$ relation. The feedback from AGNs seems to be an important component for the co-evolution of SMBHs and galaxies. Outflows generated by AGNs can shut off star formation in a `negative' feedback \citep{fab12}. On the other hand, spatially resolved spectroscopic observations revealed that, in certain circumstances, AGN outflows can also trigger star formation in a `positive' feedback \citep{mai17,gal19}. This, again, confirms that the role of the AGN in the possible co-evolution of SMBHs and the host galaxies is very complex. The \divingTD data will allow us to analyse, with high spatial and spectral resolution, the gas kinematics in the surroundings of the AGNs. We will be able to detect outflows and to determine connections between them and the star formation in these central regions of the galaxies (e.g. \citealt{din17,dah19b,das20b}). The analysis of the stellar kinematics will reveal possible rotating stellar disks around the central SMBHs and, in certain cases, will allow us to estimate the masses of the central SMBHs (e.g. \citealt{men15b,men18b,ric20}), which will result in contributions to studies related to the $M$--$\sigma$ relation.

The \divingTD may also shed a light on the census of LLAGNs. Most of the LLAGNs are LINERs \citep{ho08,hec14} and their statistics is limited by the sensitivity of the detection techniques and by the spatial resolution of the observations, which is important in order to isolate the nuclear region from its surroundings. Some attempts to detect AGNs using IFU surveys, like MaNGA, were presented in the literature (see e.g. \citealt{wyl18}); however, the fraction of nuclear activity that is detected may be quite low, since the nebular emission that is powered by AGNs is diluted by the emission that is caused by other ionization processes, mainly because of the lower spatial resolution of the observations \citep{san20}. We believe that, with the techniques developed by our group \citep{ste09,men12T,men14,men15,men19}, together with the high spatial resolution of the \divingTD data, we will be able to detect AGNs at lower luminosity limits than the current level of detection. This belief is based on two facts: first, our preliminary results (see Section~\ref{sec4}) indicate that we found a significantly higher number of objects with a broad H$\alpha$ component than anticipated from the Palomar survey. A second and perhaps more important argument comes from the \textbf{$\log N$ -- $\log S$} analysis of AGNs present in Sc--Sd galaxies. There is a very strong tendency of such objects to appear in the nearest galaxies only. In X-rays, a larger proportion than expected has been detected, but the emission from X-ray binaries may affect the statistical analysis of LLAGNs \citep{des09}. More reliable is the detection of [Ne \textsc{v}] at 14 and 24 $\mu$m (MIR) \citep{sat08} that suggests that the presence of AGNs in late-type galaxies could be possibly 4 times higher than inferred at optical wavelengths.

LINERs also pose interesting questions. If star formation and AGNs are indeed interconnected (possibly with a time delay due to AGN feedback quenching star formation), then the fact that LINERs reside among old stars poses a puzzle. Maybe the once young and luminous stars present in an earlier Seyfert phase dim to a level where they can no longer be detected in contrast to the much brighter bulge population, especially when observed through large apertures. Again, the spatial resolution of the \divingTD data, coupled to our sophisticated analysis techniques, will help identifying stellar population variations. Based on SDSS data, \citet{cid11} proposed that LINERs containing true AGNs show some residual level of recent star formation in the last Gyr, while those LINERs where stars are all old are not truly AGNs, but retired galaxies \citep{sta08}, where the ionizing photon budget is dominated not by an AGN but HOLMES. With a much greater sensitivity to AGN signatures, the \divingTD survey will help disentangling true from fake AGNs.

The \divingTD survey will certainly contribute to studies related to the AGN luminosity function. \citet{ho08} obtained a luminosity function for the H$\alpha$ emission line in AGNs, in the local universe, in the form of $\Phi \propto L^{-1.2 \pm 0.2}$. One of the sources of uncertainties for this luminosity function is the difficulty for differentiating true AGNs from impostors (LINERs powered by HOLMES and not by AGNs, for example). With the high spatial resolution of the \divingTD survey data, together with the benefits provided by our treatment and analysis techniques, we will be able to detect true AGNs in a more reliable way. To confirm the presence of an AGN in a given galaxy, we will first analyse its nuclear emission-line spectrum. If the emission-line ratios of this spectrum indicate a Seyfert, then we may confirm a true AGN. If the line ratios indicate a LINER, then we will search for a broad component in H$\alpha$. If a broad component is present, we will check if it really comes from an unresolved source by making an image of the wing (usually the red wing) of the broad component. If no broad component is detected in the spectrum, then we will need to search for other indicators of nuclear activity, such as the existence of X-ray or radio cores, or the presence of high-ionization lines in MIR spectra (e.g. the [O IV] and [Ne V] lines). Due to the selection criteria of our sample, only the AGN luminosity function in the local universe will be evaluated. Even so, the results may provide relevant information about the topic of the AGN evolution with $z$. It is currently accepted that the spatial density of more luminous AGNs increases with $z$ and peaks at $z \sim 2$. This phenomenon is usually called AGN downsizing. For $z > 2$, the density of more luminous AGNs tends to slowly decrease \citep{mer13}. The discussion of such a scenario depends also on a precise description of the AGN luminosity function at $z \sim 0$, for which the \divingTD survey will be able to contribute.

One additional relevant topic for the \divingTD survey is related to the so-called transition objects, which are usually interpreted as LINERs contaminated by the emission from \hii regions \citep{ho93, ho03}. If that is the case, one would expect that observations with sufficiently high spatial resolution would be able to disentangle the emission from the central LINER from the putative circumnuclear \hii regions and, as a consequence, the number of detected transition objects would decrease substantially. However, there is no consensus in the literature about this scenario for transition objects. \citet{shi07}, for example, analised high spatial resolution Space Telescope Imaging Spectrograph (STIS) data of 23 galaxies from the sample of the PALOMAR survey and did not find complete support for such a scenario. The \divingTD survey is intended to make a significant contribution to this topic, considering the high spatial resolution of the data.

\subsection{The main \divingTD objectives}\label{sec22}
Our main goals with this project are:

1 – \textbf{Nuclear emission-line properties.} We aim to quantify, with high signal/noise and high (seeing-limited) spatial resolution, the properties of the nuclei of all galaxies in the sample. Our objective is to detect the faintest AGNs that can currently be observed in the optical. The diagnostic diagrams of these objects may show a decrease in the number of transition objects population as \hii regions tend to be separated from LINERs/Seyferts. 

2 – \textbf{Circumnuclear emission-line properties.} The circumnuclear emission can reveal aspects about ionizing sources that may be important within a scale of $\sim$ 100 pc. Is the AGN responsible for the circumnulear emission or are there other sources of excitation/ionization such as shock waves or HOLMES? We intend to identify structures, such as ionization cones, which should allow a test of whether the unified model \citep{ant93,urr95} applies to LLAGNs as well.

3 – \textbf{Central stellar and gas kinematics.} We will obtain not only information about rotating stellar and gas discs, but also about deviations from these patterns, such as counter-rotations, kinematically decoupled cores, gas inflows and outflows, etc. Whenever possible, we intend to estimate the masses of the central SMBHs and also the mass-to-light ratio ($M/L$), using gas and stellar kinematic parameters, which may be relevant for studies related to the $M$--$\sigma$ relation \citep{fer00,geb00,gul09}.

4 – \textbf{Central stellar archaeology.} The spectral synthesis will always be attempted in order to perform the subtraction of the stellar continuum from the \divingTD data cubes. This is an essential step for a reliable analysis of the emission-line spectra of the objects. As a by-product, one may obtain information about the star-formation history (SFH) of the nucleus and of the circumnuclear environment.

\section{Methodologies for the reduction, treatment and analysis of the data}\label{sec3}

\subsection{Data reduction and data treatment}\label{sec31}

The GMOS/IFU observations of ETGs, with morphological types E, S0 - Sb, were taken with the B600 grating, which provided a wavelength coverage of 4250 - 7000 \AA ~and a spectral resolution of 1.8 \AA, measured with the sky emission line of [O \textsc{i}]$\lambda$5577. On the other hand, late-type galaxies, with morphological types Sbc - Sd, were observed with the R831 grating, resulting in a wavelength coverage of 4800 - 6890 \AA ~and in a spectral resolution of 1.3 \AA, measured with the sky emission line of [O \textsc{i}]$\lambda$5577. We opted to use the R831 grating for the observations of late-type galaxies due to their lower masses and, therefore, their lower values of the stellar velocity dispersion. As a consequence, a grating with a higher spectral resolution, such as R831, would allow a more accurate analysis of the stellar kinematics in these objects. The analysis of the stellar kinematics in higher mass ETGs, on the other hand, does not require such a high spectral resolution. Therefore, in these cases, we opted to use the B600 grating, with a longer wavelength coverage, which resulted in more accurate results obtained with stellar archaeology (see Section~\ref{sec34}).

For most of the objects, three exposures, with a spatial dithering of 0.2 arcsec per dither step, were taken, centred on the peak of the stellar emission of the galaxies. For certain objects, just one exposure, without spatial dithering, was taken. The GMOS/IFU data were reduced using the Gemini package, in IRAF environment. Bias, GCAL-flat, twilight-flat and CuAr lamp calibration images were obtained for each galaxy. A spectrophotometric standard star was also observed for each observing programme. The first step of the data reduction consisted of a simple trimming and overscan and bias subtraction. We also applied the L. A. Cosmic routine \citep{van01} for the removal of cosmic rays from the data. Then, a correction of bad pixels was performed, using a bad pixel mask obtained from the GCAL-flat images, and the spectra were extracted. Response maps obtained from the twilight-flat images were used to correct for gain variations between the fibres. Response curves provided by the GCAL-flat images were also used to correct for gain variations between the spectral pixels. The next step in the data reduction was the wavelength calibration, using the CuAr lamp images. The sky emission was subtracted using the average spectrum of the sky FOV. Since the \divingTD sample includes nearby galaxies, there was a possibility of the sky FOV in the GMOS/IFU (located at a projected distance of 1\arcmin~from the science FOV) be contaminated by the galaxy emission. In order to avoid this problem, we performed a cross-correlation between the science spectra and the sky spectra for each object in the sample, searching for spectral features of the observed galaxy in the sky FOV. For all the galaxies in the sample observed with GMOS/IFU, such features were not detected. After the sky subtraction, the data were flux calibrated, taking into account the atmospheric extinction, using the observed spectrophotometric standard stars. Finally, the last step was the construction of data cubes, with spaxels of 0.05 arcsec.

The SIFS observations were taken with a grating of 700 l/mm, providing a spectral coverage of 4500 - 7300 \AA ~and a spectral resolution of 1.3 \AA, measured with the sky emission line of [O \textsc{i}]$\lambda$5577. Six exposures, with a spatial dithering of 0.3 arcsec per dither step, were obtained for the central region of the galaxy. The data reduction was performed using scripts in Interactive Data Language (IDL). Bias, flat, milky-flat, HgAr and spectrophotometric standard stars calibration images were obtained for the observing night. Similarly to the case of GMOS/IFU data, the first step of the SIFS data reduction was a simple trimming and bias and overscan subtraction. The science image was then divided by the normalized milky-flat image (which is essentially a flat-field image with a degraded focus), in order to correct for gain variations between pixels. After that, the spectra were extracted and, using the flat image, a correction for gain variations between the fibres was applied. The data were wavelength calibrated, using the HgAr images. SIFS has also a FOV, separated from the science FOV, for the observation of the sky emission. The average spectrum of such a sky FOV was subtracted from the science spectra. We used the cross-correlation technique described before to check for possible contaminations of the sky spectra by the galaxy emission. However, no such a contamination was detected so far in the objects in the \divingTD sample observed with SIFS. Finally, the data were flux calibrated, taking into account the atmospheric extinction, using the spectrophotometric standard star observed at the same night, and data cubes were constructed, with spaxels of 0.3 arcsec.

After the data reduction, a data treatment was applied to all data cubes. The treatment procedure used for the GMOS/IFU data cubes is described in detail in \citet{men19} and included: correction of the differential atmospheric refraction (DAR); combination of the data cubes of each galaxy into one in the form of a median, to remove remaining cosmic rays and bad pixels not removed during the data reduction; Butterworth spatial filtering, to remove high spatial frequency noise from the images of the data cubes; ``instrumental fingerprint'' removal, which is essential, as this instrumental feature (usually presenting a low-frequency spectral signature and appearing as horizontal or vertical stripes across the images) can significantly compromise the analyses to be performed; and Richardson-Lucy deconvolution, to improve the spatial resolution of the data cubes. In the case of SIFS data cubes, the treatment procedure was similar, except that the DAR correction was not applied because SOAR has an Atmospheric Dispersion Corrector and, as a consequence, the obtained data cubes do not show the DAR effect. For more detail about the data treatment procedure, see \citet{men14,men15}. Fig.~\ref{fig04} shows, as an example, the image of the GMOS/IFU data cube of the central region of the galaxy NGC 5236, collapsed along the spectral axis, together with its average spectrum. Fig.~\ref{fig05} shows the same, but for the SIFS data cube of the galaxy NGC 1448.

\subsection{Analysis of the emission-line spectra}\label{sec32}

For a reliable analysis of the emission-line spectra of the data cubes, we first apply a subtraction of the stellar continuum using synthetic stellar spectra provided by the Penalized Pixel Fitting (pPXF) procedure \citep{cap04}, which is performed by fitting the observed stellar spectra with combinations of template stellar population spectra, from a given base, convolved with a Gauss-Hermite expansion. Fig.~\ref{fig06} shows the fit provided by the pPXF technique applied to a spectrum extracted from the data cube of NGC 5236. The fit residuals, which represent the emission-line spectrum, are also shown.

Emission lines are analysed in terms of Gaussian decomposition. Normally, this procedure consists of fitting the [S \textsc{ii}] and the [N \textsc{ii}]+H$\alpha$ emission lines with a two-kinematic component model. However, in certain cases, a single-kinematic component model is used. In order to reproduce a possible broad component of the H$\alpha$ emission line, a broad Gaussian function is added to the model, when necessary. By applying this process, we are able to determine the flux ratios of the narrow components of different emission lines. Diagnostic diagrams \citep{bal81,vei87} are used to classify the emission-line spectra of the data cubes as being characteristic of LINERs, Seyferts, transition objects or \hii regions. Such a procedure is applied to both nuclear and circumnuclear regions of the galaxies. Fig.~\ref{fig07} shows the Gaussian fits applied to the [S \textsc{ii}] and [N \textsc{ii}]+H$\alpha$ emission lines of the nuclear spectrum of NGC 4594.

\subsection{Analysis of the stellar and gas kinematics}\label{sec33}

The pPXF technique also provides parameters related to the stellar kinematics, such as the stellar radial velocity ($V_*$), the stellar velocity dispersion ($\sigma_*$) and the Gauss-Hermite coefficients. Since this procedure is applied to the spectrum corresponding to each spaxel of the data cubes, we obtain maps of all these kinematic parameters. Fig.~\ref{fig08} shows the maps of the stellar kinematic parameters of the data cube of the galaxy NGC 1395. The gas kinematic parameters are obtained by fitting Gaussian functions to the main emission lines of the spectra, such as the [N \textsc{ii}]+H$\alpha$ emission lines. Again, since the Gaussian fits are applied to the spectrum corresponding to each spaxel of the data cubes, the results are maps of the gas kinematic parameters. Fig.~\ref{fig09} shows maps of the gas radial velocity ($V_\mathrm{gas}$) and of the gas velocity dispersion ($\sigma_\mathrm{gas}$) of the data cube of NGC 1395.

\begin{figure*}
 \centering 
 \includegraphics[width=\textwidth]{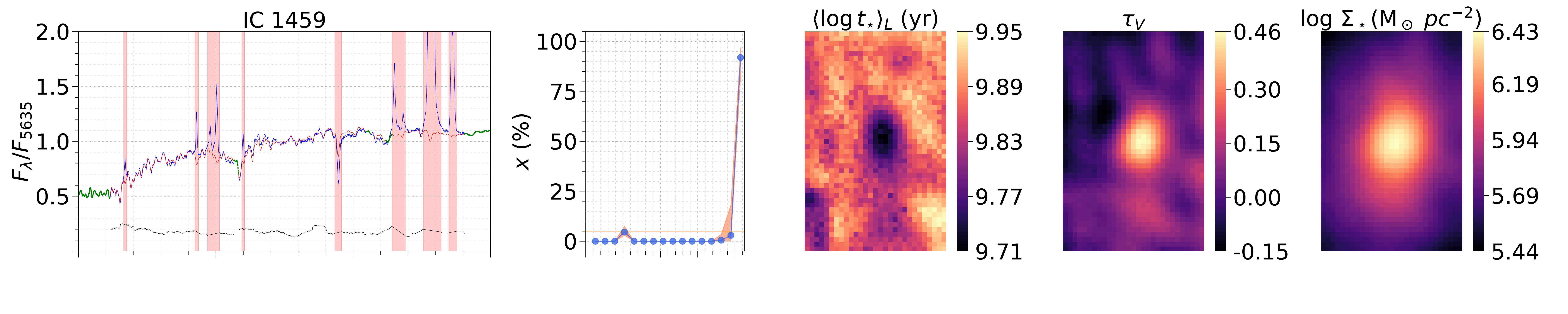}\vspace*{-5mm}
 \includegraphics[width=\textwidth]{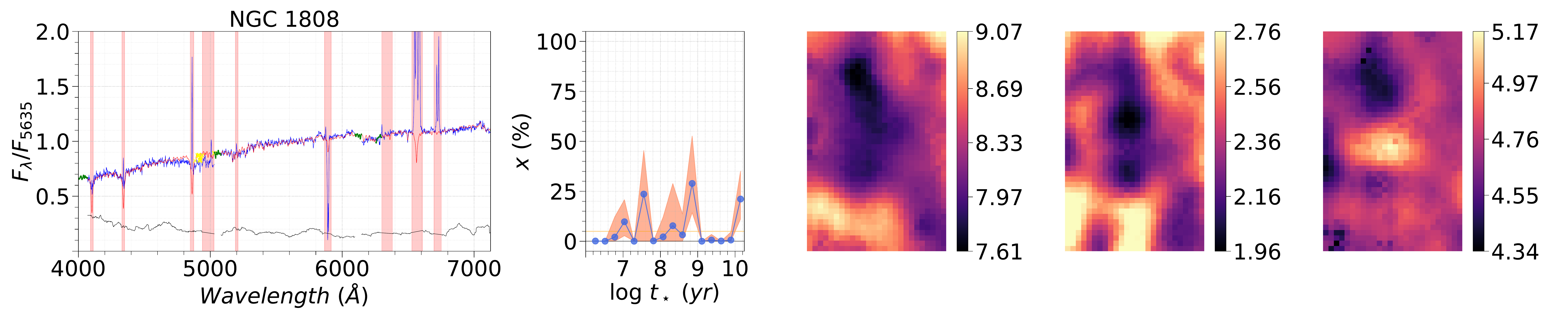} 
 \caption{Stellar population properties derived with {\sc starlight} for the ellipical galaxy IC 1459 ({\it top}) and the spiral galaxy NGC 1808 ({\it bottom}).  
 The first panel shows the observed spectrum (blue), the best-fitting model (red) and the error spectrum (black). Green portions of the spectrum indicate regions without data or bad pixels and red areas are masked regions. Second panel is the SFH. The blue curve is the mean light fraction in each base component measured over the data cube. The areas in red indicate the 10-95th percentile range and the horizontal orange line is a visual reference for 5 per cent of the light fraction. The following three panels are the maps of light-weighted mean age, extinction and stellar mass surface density, respectively.
}  
 \label{fig10}
\end{figure*}

\subsection{Stellar archaeology}\label{sec34}

The stellar archaeology parameters can be derived by means of the \textsc{starlight} spectral synthesis code \citep{cid05}, similarly to what was done for galaxy-wide data cubes in the CALIFA project \citep[e.g.,][]{per13,cid13,gon14}.
Fig. \ref{fig10} shows example maps of light-weighted mean age, dust optical depth and stellar mass surface density for the elliptical galaxy IC 1459 (top panels) and for the spiral galaxy NGC 1808 (bottom). The plot also shows the spectral fits and the derived light fractions (at 5635 \AA) associated with populations of different ages.

One should note that the spectral coverage of \divingTD does not include the blue/near-ultraviolet region of the spectrum, which contains important indicators of the age and metallicity of stellar populations \citep{leo96,wor97,tra98}. This limitation translates to difficulties and possibly biases in the inferred stellar population properties.
Indeed, experiments carried out by \citet{car19} show that the light-weighted mean ages derived from data in the 4500-7000\,\AA\ spectral range tend to be younger than those obtained using a wider 3700-7000\,\AA\  window -- see also \citet{wol07},  \citet{cid14} and \citet{Goncalves+2020} for related experiments. 

Nuclear activity itself is another complicating factor for stellar archaeology work. In some cases the amount of emission lines (which are masked from the spectral fits) in  the nuclear region leave little room for a reliable stellar population analysis, as {\sc starlight} has few or no strong absorption lines to anchor the fit. Furthermore, one should also allow for the presence of a non-stellar continuum. NGC 1566 is a good example of these issues. The analysis of the \divingTD data cube of this galaxy by \citet{das17} found a $f_\nu \propto \nu^{-1.7}$ in the nuclear region (r < 0.5 arcsec).

Despite these caveats, our previous experience (e.g.\ \citealt{das17,das18, men18,das20b}) show that this process can, at least, differentiate, with good precision, emission from young and old stars and also detect differences in the metallicities of the stellar populations.

\section{Early results}\label{sec4}

To date 20 papers showing specific analyses of \divingTD data have already been published. Studies (focused on the line-emitting regions, the stellar and gas kinematics and the stellar archaeology) of the central regions of six individual galaxies of the \divingTD sample resulted in the following publications: NGC 4594 \citep{men13}, NGC 1313 \citep{men17}, NGC 5044 \citep{din17}, NGC 7582 \citep{ric18}, NGC 1052 \citep{dah19a,dah19b} and NGC 2835 \citep{men19}. In certain objects, peculiar features were detected, such as an off-centred AGN in NGC 3115 \citep{men14b}, a compact off-centred Seyfert emission (with a deficit of emission at the AGN position) in NGC 3621 \citep{men16}, double stellar nuclei in NGC 908 and NGC 1187 \citep{men18} and a double-peaked H$\alpha$ emission in NGC 4958 \citep{ric19}.

One of the subsamples of the \divingTD includes the Milky Way morphological twins \citep{das16M,das20T}. Five papers about the analyses of individual objects of such a subsample were already published. The first \citep{das17} was focused on the Seyfert 1 nucleus of NGC 1566. The second \citep{das18} presented the analysis of the central region of NGC 6744, a nucleus with old stellar populations and LINER emission. The other three papers described the rich environments in the central regions, with LINER emission, of NGC 613 \citep{das20a,das20b} and NGC 2442 \citep{das21}.

Five ETGs that belong to the \divingTD survey were part of a sample of 10 massive ETGs from the local Universe, all observed with GMOS/IFU using the same setup as the \divingTD objects \citep{ric13T, ric14a}. LINER-like nuclei were detected in all 10 objects \citep{ric14b}, with six of them being type-1 AGNs. In addition, \citet{ric15} studied the properties of the extended emission of these objects and proposed a model for LINER-like circumnuclear regions where the emission along gas disks are caused by an extended-ionization source, probably HOLMES, while the emission in the direction perpendicular to the disks are related to a low-velocity ionization cone where the ionizing photons from the AGN are collimated by some agent aligned with the gas disk. Finally, \citet{ric16} studied the stellar kinematic properties of this sample. Seven galaxies have a rotation structure within their central region, i.e. in a scale of $\sim$ 100 pc. In NGC 1404, which is part of the \divingTD survey, a kinematic decoupled core was clearly detected. This series of papers will serve as a guide to the analysis that will be performed in the ETG subsample of the \divingTD survey.   

Regarding the statistical results, a preliminary analysis of the subsample of ETGs in the \divingTD survey revealed that 91 per cent of these objects show nuclear emission lines \citep{ric21iau}. In addition, the nuclear emission-line spectra of 50 per cent of the ETGs are typical of LINERs and Seyferts (31 per cent are actually type 1 AGNs). The analysis of the subsample of galaxies brighter than B=11.2 (which we call mini-\divingTD sample; \citealt{men21iau}, Menezes et al. in preparation) reveals similar results, with 93 per cent of the objects with nuclear emission lines and 50 per cent showing emission-line spectra typical of LINERs or Seyferts. One interesting point is that the diagnostic diagram analysis of the mini-\divingTD sample indicates an apparent dichotomy between \hii regions and LINERs/Seyferts, with few transition objects. This suggests that at least part of the transition objects are LINERs contaminated by the emission from \hii regions and the high spatial resolution of the \divingTD data allows a more accurate separation between the nuclear and circumnuclear emission, resulting in a lower number of transition objects.

\section{Summary}\label{sec5}

We are conducting the \divingTD survey, which has the goal of analysing, using optical 3D spectroscopy, the central regions of all galaxies in the Southern hemisphere with $B < 12.0$ and $|b| > 15\degr$. The complete sample has a total of 170 objects. From this sample, 131 galaxies were observed with the GMOS/IFU, at the Gemini South telescope, eight were observed also with GMOS/IFU but at the Gemini North telescope and 25 were observed with SIFS, at the SOAR telescope. It is intended that the remaining six galaxies will be observed with SIFS in the near future. The data obtained for this survey has a combination of high spatial and spectral resolutions not matched by previous surveys and will result in significant contributions to different research areas, such as the statistics of LLAGNs for galaxies with different morphological types, the ionization mechanisms in LINERs, the nature of transition objects, etc. The main objectives of the \divingTD survey are the study of: nuclear emission-line properties; circumnuclear emission-line properties; central stellar and gas kinematics; and central stellar archaeology. 

The survey already resulted in 20 published papers, showing the analyses of the central regions of individual objects or small subsamples of the \divingTD survey. The first statistical results of the \divingTD survey, which are based on the analysis of the nuclear region of all galaxies of the sample with B $<$ 11.2 (the mini-\divingTD sample), are presented in the second paper of this series (Menezes et al, submitted, but see also \citealt{men21iau} for preliminary results on this sample). A summary containing preliminary results of the nuclear region of all ETGs of the \divingTD survey is presented in \citet{ric21iau}; a more complete analysis on the nuclear emission from the ETGs is in preparation. A statistical study of the Milky Way morphological twins of the sample is expected to be published in the near future. When all the objects of the \divingTD survey are observed, we intend to perform statistical studies of the complete sample and also of other subsamples, including, for example, the early spiral galaxies (Sa+Sab+Sb+Sbc) or the late spiral galaxies (Sc+Scd+Sd).

\section*{Acknowledgements}

This work is dedicated to the memory of the founder of this survey, Prof. Dr. Jo\~ao Evangelista Steiner, who sadly passed away on 2020 September 10.

Based on observations obtained at the Gemini Observatory (processed using the Gemini IRAF package), which is operated by the Association of Universities for Research in Astronomy, Inc., under a cooperative agreement with the NSF on behalf of the Gemini partnership: the National Science Foundation (United States), the National Research Council (Canada), CONICYT (Chile), the Australian Research Council (Australia), Minist\'{e}rio da Ci\^{e}ncia, Tecnologia e Inova\c{c}\~{a}o (Brazil) and Ministerio de Ciencia, Tecnolog\'{i}a e Innovaci\'{o}n Productiva (Argentina). This research has made use of the NASA/IPAC Extragalactic Database (NED), which is operated by the Jet Propulsion Laboratory, California Institute of Technology, under contract with the National Aeronautics and Space Administration. This research has also made use of the Carnegie-Irvine Galaxy Survey (https://cgs.obs.carnegiescience.edu/CGS/Home.html), the Hyperleda (http://leda.univ-lyon1.fr/) and RC3 (https://heasarc.gsfc.nasa.gov/W3Browse/all/rc3.html) databases. We thank Conselho Nacional de Desenvolvimento Cient\'ifico e Tecnol\'ogico (CNPq) for support under grants 306063/2019-0 (RBM), 306790/2019-0 (TVR) and 141766/2016-6 (PS) and Funda\c{c}\~ao de Amparo \`a Pesquisa do Estado de S\~ao Paulo (FAPESP) - for support under grants 2011/51680-6 and 2020/13315-3 (PS). NVA acknowledges support of the Royal Society and the Newton Fund via the award of a Royal Society--Newton Advanced Fellowship (grant NAF\textbackslash{}R1\textbackslash{}180403), and of Funda\c{c}\~ao de Amparo \`a Pesquisa e Inova\c{c}\~ao de Santa Catarina (FAPESC). We also thank an anonymous referee for valuable comments about this paper.  

\section*{Data Availability}

Further detail about the \divingTD survey can be found at https://diving3d.maua.br. The raw GMOS/IFU data are available at the Gemini Science Archive (https://archive.gemini.edu/searchform). The treated GMOS/IFU and SIFS data cubes can be requested at diving3d@gmail.com.



\bibliographystyle{mnras}
\bibliography{references} 





\bsp	
\label{lastpage}
\end{document}